\documentstyle[prl,aps,epsf,floats]{revtex}

\begin{document}

\twocolumn[\hsize\textwidth\columnwidth\hsize\csname @twocolumnfalse\endcsname

\author{A.M. Bratkovsky$^{1}$ and A.P. Levanyuk$^{1,2}$}

\address{$^{1}$ Hewlett-Packard Laboratories, 1501 Page Mill Road, Palo
Alto, California 94304\\
$^{2}$Departamento de F\'{i}sica de la Materia Condensada, C-III,
Universidad Aut\'{o}noma de Madrid, 28049 Madrid, Spain
}
\title{
Smearing of phase transition 
due to a surface effect or a bulk 
inhomogeneity in ferroelectric nanostructures
}
\date{January 30, 2004}
\maketitle

\begin{abstract}

The boundary conditions, customarily used in the 
Landau-type approach to ferroelectric thin films and nanostructures,
have to be modified to take into account that 
a surface of a ferroelectric (FE) is a defect of the ``field'' type.
The surface (interface) field  is coupled to a normal component of polarization
and,  as a result, the second order phase transitions 
are generally suppressed and anomalies in response 
are washed out.
In FE films with a compositional (grading) or some other type of
inhomogeneity, the transition 
into a monodomain state is suppressed, but 
a transition with formation of a domain structure may occur.

\end{abstract}

\vskip2pc]

\narrowtext

Theoretical studies of phase transitions in thin films and the corresponding
size effects within the Landau theory \cite{Landau37a,Landau37b} have been
undertaken since 1950s. Recently the interest to these questions has risen
dramatically in view of the applications of ferroelectric thin films \cite
{scott00} and a discovery of various ferroelectric nanostructures\cite
{nanofe}. The boundary conditions for thin films were originally discussed
by Ginzburg and Landau (GL) in 1950\cite{GinzburgLandau} and by Ginzburg and
Pitaevskii in 1958 \cite{GPit58}. It was shown by GL that, if the properties
of the boundary layer are the same as of the bulk, one arrives at the
condition that the gradient of the order parameter vanishes at the surface, $%
\vec{\nabla}_{\vec{n}}\eta =0$ (in zero magnetic field, $\vec{n}$ is the
normal to the surface). Starting from a microscopic theory, de Gennes has
shown that for a superconductor-metal interface with no current and magnetic
field a more general boundary condition applies, $\vec{\nabla}_{\vec{n}}\eta
+\eta /\delta =0,$ where $\delta $ is the characteristic length scale
describing the proximity effect \cite{degennes}. These conditions are very
general and were obtained phenomenologically by Kaganov and Omelyanchouk for
a surface of a ferromagnet \cite{Kaganov} (cf. review in \cite{Binder72}).
Kretschmer and Binder \cite{Binder79}, using the same boundary conditions,
have taken into account the depolarizing field, which is important when a
ferroelectric polarization (or magnetization) is perpendicular to the
surface. Later these boundary conditions have been used customarily in
studies of phase transitions in ferroelectric films (see, e.g. \cite
{Tilley2001}).

It is obvious, however, that while the treatment \cite{Kaganov,Binder79} is
appropriate for magnetics, it overlooks an important specific feature of
ferroelectric phase transitions in thin films, wires, and other systems with
boundaries. Indeed, there is an effective field at the surface (interface)\
of any material appearing because the surface breaks the symmetry of the
bulk. For instance, a part of this surface field might be due to a Coulomb
dipole field (double layer), contributing to the work function \cite
{bardeen36,Monch}. This makes ferroelectric surfaces {\em qualitatively
different} from the surfaces of magnetics. The effective field is coupled to
the component of polarization perpendicular to the surface/interface and, as
a result, the second order ferroelectric phase transitions are typically
smeared out, as we shall see below.

We shall discuss, as an example, a paraelectric-ferroelectric phase
transition in cubic perovskite thin films where a surface is perpendicular
to one of the cubic axes. The surface or interface eliminates all the
symmetry elements, which change a vector perpendicular to the surface and
generates a local field conjugated to the polarization component
perpendicular to the surface ($P_{z}$ in our case)\cite{LevSig88}. To
demonstrate the effect, we consider the state with $P_{x},$ $P_{y}=0$,
described for given potentials on electrodes \cite{Chensky,BL} by the free
energy $\tilde{F}_{f}=F_{LGD}+\int dV\frac{E^{2}}{8\pi }-\sum_{a}e_{a}%
\varphi _{a},$ 
\begin{equation}
F_{LGD}=\int dV\left[ \frac{1}{2}AP^{2}+\frac{1}{4}BP^{4}+\frac{1}{2}g\left( 
\frac{dP}{dz}\right) ^{2}+\frac{1}{2}D\left( \nabla _{\perp }P\right) ^{2}%
\right] ,  \label{eq:Ft}
\end{equation}
where $\nabla _{\perp }=(\partial /\partial x,\partial /\partial y)$ is the
gradient in a plane of the film, $q_{a}$ ($\varphi _{a})$ are the charges
(electrostatic potentials)\ at the electrodes $a=1,2$. Here $\nabla _{\perp
}=0$\ for the monodomain state. We assume ideal electrodes with a vanishing
Thomas-Fermi screening length. As discussed above, the surface produces an
effective surface field $w,$ and, generalizing Ref.\cite{Kaganov}, we have
to add the surface energy to (\ref{eq:Ft}) to obtain the free energy of the
film 
\begin{equation}
\tilde{F}=\tilde{F}_{f}+\int dS\left( \frac{1}{2}\alpha P^{2}-wP\right) ,
\label{eq:Fws}
\end{equation}
where $\alpha $ corresponds to a ``temperature''-like component of the
surface energy. We obtain from Eqs.~(\ref{eq:Ft})\ and (\ref{eq:Fws})\ after
an integration by parts the {\em correct boundary conditions for
ferroelectrics} 
\begin{equation}
\alpha _{1(2)}P+(-)g\frac{dP}{dz}=w_{1(2)},\quad z=+(-)l/2.  \label{eq:BCsrf}
\end{equation}
One can estimate that $\alpha \sim d_{at},$ where $d_{at}$ is the
characteristic ``atomic'' length scale, on the order of the lattice
constant. The electric field at the surface, first considered many decades
ago \cite{bardeen36}, is on the order of $w/d_{at}\sim \Phi
_{s}/d_{at}\sim 1$V/$\AA\approx 10^8$V/cm, 
 where $q\Phi _{s}\sim 4$eV\ is the typical workfunction
for ferroelectrics\cite{scott00}. 
The surface bias field  corresponds to a surface
charge $\sim 100\mu C/$cm$^{2}$, which is on the
order of an ``atomic'' polarization $P_{at}=q/d_{at}^{2}\sim 200\mu C/$cm$%
^{2}$, so that $w\sim P_{at}d_{at}$ (we expect that the non-Coulomb
contribution to $w$ is of the same order of magnitude). The polarization $%
P\left( z\right) $ is found from the equation of state for (\ref{eq:Ft}) and
the Poisson equation, assuming that there is no external charge, and
neglecting for a moment the non-linear terms in polarization: 
\begin{eqnarray}
AP-g\frac{d^{2}P}{dz^{2}} &=&E,  \label{eq:Eqstate} \\
\frac{d}{dz}\left( E+4\pi P\right) &=&0,  \label{eq:Gauss} \\
\frac{1}{l}\int_{1}^{2}Edz &=&\frac{\varphi _{1}-\varphi _{2}}{l}=\frac{U}{l}%
\equiv E_{0},  \label{eq:biasv}
\end{eqnarray}
where $E_{0}$ is the external electric field. We obtain from Eqs.~(\ref
{eq:Gauss}) and (\ref{eq:biasv})

\begin{equation}
E=E_{0}-4\pi \left[ P\left( z\right) -\bar{P}\right] ,  \label{eq:field}
\end{equation}
where the overbar means an average over the film, i.e. $\ \bar{f}%
=(1/l)\int_{-l/2}^{l/2}dzf(z).$ Substituting this into Eq.~(\ref{eq:Eqstate}%
) and integrating over the film, we find 
\begin{equation}
A\bar{P}-\frac{g}{l}\left( \frac{dP(l/2)}{dz}-\frac{dP(-l/2)}{dz}\right)
=E_{0}.  \label{eq:Eqstate2}
\end{equation}
We write down the solution as a sum $P=P_{0}+p(z)$ of the homogeneous, $%
P_{0}=\left( E_{0}+4\pi \bar{P}\right) /\left( A+4\pi \right) ,$ and the
inhomogeneous, $p(z),$ term 
\begin{eqnarray}
p &=&C_{1}\exp \left[ -\lambda (z+l/2)\right] +C_{2}\exp [-\lambda (l/2-z)],
\\
C_{1(2)} &=&\left( w_{1(2)}-\alpha _{1(2)}P_{0}\right) /\left( \alpha
_{1(2)}+\lambda g\right) .  \label{eq:C12}
\end{eqnarray}
where $\lambda =\sqrt{\frac{A+4\pi }{g}}\approx \sqrt{\frac{4\pi }{g}}\sim
d_{at}^{-1},$. Since $\bar{P}=P_{0}+(C_{1}+C_{2})/\lambda l,$ we obtain with
the use of Eqs.~(\ref{eq:C12}),(\ref{eq:Eqstate2}) 
\begin{equation}
A^{\prime }\bar{P}=E^{\prime }+\frac{\lambda g}{l}\left( \frac{w_{1}}{\alpha
_{1}+\lambda g}+\frac{w_{2}}{\alpha _{2}+\lambda g}\right) ,
\label{eq:eqst4}
\end{equation}
where $A^{\prime }=A(1-\xi /\lambda l)+\lambda g\xi /l\approx A+4\pi \xi
/\lambda l,$ $E^{\prime }=E_{0}\left( 1-\xi /\lambda l\right) ,$ and $\xi
=\alpha _{1}/(\alpha _{1}+\lambda g)+\alpha _{2}/(\alpha _{2}+\lambda g)\sim
1.$ The phase transition in this case is smeared out, since generally the
surface dipoles are asymmetric$.$ In the symmetric case, $w_{1}=-w_{2},$ $%
\alpha _{1}=\alpha _{2},$ the phase transition persists, but the transition
temperature of a transition into a monodomain state is shifted by the amount
inversely proportional to the thickness of the film, with the following
estimate for displacive systems (cf. Ref.~\cite{Binder79}): 
\begin{equation}
\Delta T_{c}=\frac{4\pi \xi T_{at}}{\lambda }\frac{1}{l}.  \label{eq:dTc}
\end{equation}
The monodomain transition in the symmetric case occurs at $A=-\lambda g\xi
/l\approx -d_{at}/l.$ This is close to a transition with the formation of
domains \cite{BLinh}. Which transition actually occurs depends on materials
parameters.

The surface dipoles discussed above are a special case of polarization due
to gradients of a scalar quantity (concentration $c$, density, temperature,
etc.) and they are accounted for by a term like 
\begin{equation}
f_{c}=-\gamma \vec{P}\vec{\nabla}c,  \label{eq:fgrad}
\end{equation}
in the free energy, where the coefficient $\gamma $\ is estimated as $\gamma
\sim P_{at}d_{at}$ \cite{Kogan64,LevMin80} (see also \cite{Tagan87}).

Consider now the case of a film with a compositional profile (grading)\
given by e.g. the concentration of one of the components of a ferroelectric
alloy $c=c(z)$. Such systems are currently a focus of research in
ferroelectrics due to their unusual pyroelectric characteristics\cite{graded}%
. The equation of state of the graded ferroelectric film is 
\begin{equation}
A(z)P+BP^{3}-g\frac{d^{2}P}{dz^{2}}-D\nabla _{\perp }^{2}P=E_{0}+4\pi (\bar{P%
}-P)+\gamma \frac{dc}{dz}.  \label{eq:AzgamES}
\end{equation}

Consider a special case of a {\em step-wise concentration profile}, i.e. $%
c=c_{1}$ when $0<z<l_{1}$, and $c=c_{2}$ when $-l_{2}<z<0,$ and the boundary
conditions are ``neutral'' ($dP/dz=0$ at $z=l_{1},-l_{2}$). The equation of
state in this case is 
\begin{equation}
A_{r}P+BP^{3}-g\frac{d^{2}P}{dz^{2}}=E_{0}+4\pi (\bar{P}-P),  \label{eq:eqpw}
\end{equation}
for the both parts of the film $r=1,2.$ The boundary conditions at $z=0$
follow from the continuity of a displacement field $E+4\pi P$ and the
equation of state (\ref{eq:AzgamES}). In displacive systems the electric
field $E\sim AP$ is always much smaller than the polarization $P,$ since $%
|A|\ll 1$\cite{BLinh}. Hence, with high accuracy $\propto (c_{1}-c_{2})/4\pi
\ll 1,$ the boundary conditions are 
\begin{eqnarray}
P_{1} &=&P_{2}  \nonumber \\
g\left( \frac{dP_{1}}{dz}-\frac{dP_{2}}{dz}\right) &=&-h,  \label{eq:BCdc}
\end{eqnarray}
at $z=0,$ with $h=\gamma (c_{1}-c_{2})\sim P_{at}d_{at}(c_{1}-c_{2}).$

We have studied before \cite{BLinh} a similar situation but without the
concentration gradient, i.e. for $\gamma =0.$ Having assumed that the $z-$%
dependence of $A$ is due to the concentration dependence of the Curie
temperature with $T_{c1}-T_{c2}\sim T_{at}(c_{1}-c_{2})$ and $dA/dT\sim
T_{at}^{-1}$ (displacive systems) we have shown that for $%
(c_{1}-c_{2})\gtrsim d_{at}/l$ the loss of stability of a paraphase occurs with a
formation of a domain structure, and it takes place at $A_{2}>0$ but $%
A_{1}<0 $ and $|A_{1}|\sim (c_{1}-c_{2})\gtrsim d_{at}/l$ \cite{BLinh}. For these
threshold values of concentration inhomogeneity the result of Ref.\cite
{BLinh} stands, whereas in the case of the bulk inhomogeneity and general 
{\em nonsymmetric} boundary conditions the results are different (see below).

The main effect of the bias field is that now there is a polarization at all
temperatures, and, therefore, the phase transition into a monodomain state
is smeared out. However, a phase transition with formation of a domain
structure is still possible. To see this, we need to investigate a stability
of a monodomain solution of (\ref{eq:AzgamES}). First, we need to find{\bf \ 
}the average polarization $\bar{P}$ across the film. Integrating Eq.~(\ref
{eq:eqpw}) over the film thickness, we obtain 
\begin{eqnarray}
&&\bar{A}\bar{P}+B\bar{P}^{3}+\overline{\delta A_{r}\delta P}+B\left( 3\bar{P%
}\overline{\delta P^{2}}+\overline{\delta P^{3}}\right)  \nonumber \\
&=&E_{0}+w/l,  \label{eq:AbPbw} \\
w &=&h+w_{1}+w_{2}-\alpha _{1}P(l_{1})-\alpha _{2}P(-l_{2}),
\end{eqnarray}
where $\bar{A}=\nu _{1}A_{1}+\nu _{2}A_{2},$ $\delta A_{r}=A_{r}-\bar{A},$ $%
\nu _{1(2)}=l_{1(2)}/l,$ $l=l_{1}+l_{2},$ Fig.~1. There are two
possibilities:\ (i)\ near {\em symmetric}, $|w_{1}+w_{2}|\lesssim |h|$ and
(ii)\ {\em asymmetric} surfaces, $\left| w_{1}+w_{2}\right| \gg |h|.$ In the
first case the monodomain transition is smeared out by the presence of the
gradient dipole field $h$, but the transition with domain formation is
possible. In the second, more general, case the monodomain transition is
smeared out, and domains either form or they do not, depending on the
concentration gradient and/or the thickness of the film. 

Importantly, the effective bias field $w/l$, conjugated to the average order
parameter $\bar P$, Eq.~(\ref{eq:AbPbw}), is large.
For a  1000$\AA$ thick film it would have the same effect as if there were
an external electric field $\sim 10^5$~V/$\AA$, which, for comparison,
is only marginally smaller than the breakdown field in graded FE $E_b
\approx 0.75 \cdot 10^6$~V/cm\cite{graded}. 

\begin{figure}[t]
\epsfxsize=2.8in \epsffile{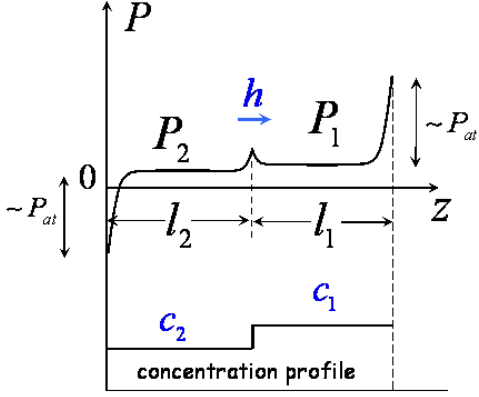}
\caption{ Schematic of the polarization distribution in a ferroelectric with
a step-wise concentration profile.}
\label{fig:PzDc}
\end{figure}
To investigate a stability loss of a paraphase, we assume that the linear
approximation is valid and check later if the solution justifies the
assumption. We estimate $\bar{P}\approx w/\bar{A}l\sim
P_{at}d_{at}/(c_{1}-c_{2})l,$ hence the first term in (\ref{eq:AbPbw}) is $\bar{A}\bar{P%
}\sim w/l\sim P_{at}d_{at}/l.$ The cubic term is $B\bar{P}^{3}\sim
P_{at}^{-2}\left( w/\bar{A}l\right) ^{3}\sim P_{at}\left( d_{at}/\bar{A}l\right)
^{3},$ since $w\sim P_{at}d_{at},$ and we can neglect it in comparison with
the first linear term in (\ref{eq:AbPbw}) when $\bar{A}\sim
(c_{1}-c_{2})\gtrsim \left( d_{at}/l\right) ^{2/3}$. The latter is the condition
for the {\em linearization} of the equation of state (\ref{eq:AzgamES}),
which takes the form 
\begin{equation}
\left( \tilde{A}_{r}+4\pi \right) \delta P-gd^{2}\delta P/dz^{2}=-w/l-\delta
A_{r}\bar{P},  \label{eq:linw}
\end{equation}
with a solution 
\begin{eqnarray}
\delta P_{r}(z) &=&-\frac{w/l+\delta A_{r}\bar{P}}{\tilde{A}_{r}+4\pi }%
+p_{r}(z),  
\label{eq:dPw} \\
p_{1} &=&ae^{-\lambda _{1}z}+C_{1}e^{-\lambda _{1}(l_{1}-z)},\quad 0<z<l_{1},
\label{eq:p1}\\
p_{2} &=&be^{\lambda _{2}z}+C_{2}e^{-\lambda _{1}(z+l_{2})},\quad
-l_{2}<z<0.
\label{eq:p2}
\end{eqnarray}
We can replace $\lambda _{1}=\lambda _{2}=\lambda =\sqrt{4\pi /g}$ and
obtain from the boundary conditions $a=b=h/(2\lambda g)$ and 
\begin{eqnarray}
\lambda gC_{1}+\alpha _{1}P(l_{1}) &=&w_{1},  \nonumber \\
\lambda gC_{2}+\alpha _{2}P(-l_{2}) &=&w_{2}.
\end{eqnarray}
A reasonable approximation is $P(l_{1})\approx
C_{1}=w_{1}/\left( \alpha _{1}+\lambda g\right) ,$ $P(-l_{2})\approx
C_{2}=w_{2}/\left( \alpha _{2}+\lambda g\right) ,$ Fig.~1. From
first-principles calculations at the surfaces of BaTiO$_{3}$ and PbTiO$_{3}$ 
$P\sim 10^{-1}P_{at}$ (see, e.g. \cite{vbilt01}).

Consider the third and fourth terms on the left hand side of Eq.~(\ref
{eq:AbPbw}).  With the use of Eqs.~(\ref{eq:dPw})-(\ref{eq:p2}) we
obtain the estimate 
$\overline{\delta
A_{r}\delta P}\sim (c_{1}-c_{2})w/4\pi l\ll \bar{A}\bar{P},$ since there is
an additional small factor $(c_{1}-c_{2})/4\pi .$ 
Both terms in $\delta P_{r}(z)$ give contributions to this estimate of
the same order of magnitude.
The term $B\bar{P}\overline{\delta
P^{2}}\sim P_{at}^{-2}\frac{w}{\bar{A}l}\left( \frac{w}{4\pi l}\right) ^{2}.$
The condition that it is smaller than $\bar{A}\bar{P}$ reads $\bar{A}\sim
(c_{1}-c_{2})\gg (d_{at}/4\pi l)^{2}$, and it is certainly obeyed when $%
(c_{1}-c_{2})\gtrsim \left( d_{at}/l\right) ^{2/3},$ which is the condition for
the linearization, obtained above. The last term in (\ref{eq:AbPbw}) is very
small if$\ (d_{at}/4\pi l)^{2}\ll 1,$ which is always the case.

Now, we shall see if the domain formation is possible. Following the
procedure of Ref.\cite{BLinh}, we have to linearize (\ref{eq:AzgamES}) about
the monodomain solution (inhomogeneous along $z$ direction only) and look
for its non-trivial solutions in the ``soft'' part of the film with $A_{1}<0$
in the form of the ``polarization wave'', $P(x,z)=\bar{P}+\delta P(z)+\xi
(x,z),$ where $\xi (x,z)\propto e^{ikx}.$ We arrive at the same problem as
in Ref. \cite{BLinh} but with a renormalized coefficient $A_{1}\rightarrow 
\tilde{A}_{1}=A_{1}+3B\bar{P}^{2}$. The boundary conditions for $\xi (x,z)$
are exactly the same as in Ref.\cite{BLinh}, and $A_{1}$ enters the
condition for instability ($A_{2}>0$ for the ``hard'' part does not). Domain
formation is possible when $\tilde{A}_{1}\sim -d_{at}/l,$ in spite of a positive
renormalization. The condition for this reads $\bar{A}\gg (d_{at}/l)^{1/2}.$ This
condition is {\em stricter} than the one for the linearity of the equation
of state (\ref{eq:AbPbw}), meaning that our using of the linearized equation
for $\bar{P}$ is justified. There is also a range of concentration
gradients, $(d_{at}/l)^{2/3}<\bar{A}\sim (c_{1}-c_{2})<(d_{at}/l)^{1/2},$ when one can
linearize the equation of state for $\bar{P}$, but domains do not form ($%
\tilde{A}_{1}>0$). Finally, when the concentration gradient is even smaller, 
$\bar{A}<(d_{at}/l)^{2/3},$ there is no domain formation and the equation for $%
\bar{P}$ is substantially nonlinear, $\bar{A}\bar{P}<B\bar{P}^{3}$.\
Therefore, the phase transition into monodomain state is smeared out, but a
phase transition with the {\em domain formation} occurs when, in general,
the concentration gradient is large enough, $c_{1}-c_{2}>(d_{at}/l)^{1/2},$ or
if, for a given concentration gradient, the film exceeds some {\em critical
thickness}, $l>l_{d}=d_{at}/\left( c_{1}-c_{2}\right) ^{2}.$ If the system does
split into domains in presence of the built-in surface bias field, the
opposite domains will have different absolute values of polarization.

In a special case of {\em symmetric} surfaces, when $|w_{1}+w_{2}|\lesssim
|h|,$ the domains {\em always} form. Here we find that the net polarization
is due mainly to concentration inhomogeneity and is much smaller than in the
general case considered above, $\bar{P}\approx h/\bar{A}l\sim P_{at}d_{at}/l.$\
This means that the term $B\bar{P}^{3}$ is on the order of $P_{at}\left(
d_{at}/l\right) ^{3}$ (since $B\sim P_{at}^{-2}$), at the same time the term $%
\bar{A}\bar{P}\sim (c_{1}-c_{2})P_{at}d_{at}/l$, i.e. for $\bar{A}\sim
(c_{1}-c_{2})\gtrsim \left( d_{at}/l\right) ^{2}$ the linear term indeed
dominates in (\ref{eq:AbPbw}), $\bar{A}\bar{P}\gg B\bar{P}^{3}.$ We obtain
also that $\delta P\sim h/4\pi l\approx \bar{A}\bar{P}/4\pi \sim
(c_{1}-c_{2})\bar{P}/4\pi \ll \bar{P}.$ Because of this relation, all terms
on the left hand side of Eq.~(\ref{eq:AbPbw}) are indeed small in comparison
with the first one, $\bar{A}\bar{P},$ and can be omitted. In the region of a
stability loss with respect to domains $A_{1}\sim -d_{at}/l$, therefore, $3B\bar{P%
}^{2}$ $\sim \left( d_{at}/l\right) ^{2},$ i.e. the positive renormalization of $%
A_{1}$ is very small, $\tilde{A}_{1}=A_{1}+3B\bar{P}^{2}\approx A_{1}<0$ and
the system splits into domains. Therefore, for symmetric surfaces{\bf \ }the
presence or absence of the interfacial bias field at the boundary between
two ferroelectric layers does not change our earlier prediction that
practically any inhomogeneity, however small, would lead to a domain
formation\cite{BLinh}.

One should make a reservation in case the boundary conditions correspond to
a ``surface ferroelectricity'' ($\alpha <0$), then the monodomain transition
can occur before a domain structure forms. However, this is a somewhat
special case and, more importantly, effect of any real electrodes is rather
opposite: it tends to suppress the ferroelectric transition into a
monodomain state \cite{BLelectrodes}. Therefore, it may be fairly difficult
to observe the effects of a\ ``surface ferroelectricity'' in the case of
spontaneous polarization {\em normal} to electrodes. Certainly, the real
boundaries are never planar but rather rough. It seems likely that in real
samples there are regions where the bias field is much smaller than would be
in the case of planar boundaries. In these ``weak'' regions even small
inhomogeneities in the materials constants would lead to a formation of
domains, just as in the case of a sample with the ``neutral'' boundary
conditions.

Note that even in a film with a step in a concentration profile the
polarization is almost constant throughout the sample (with the exclusion of
near-surface areas, Fig.~1). This is a result of a long-range depolarizing
Coulomb field. It was neglected in a recent attempt to calculate the profile
of polarization in the graded FE numerically, Ref. \cite{Alpay03}, and this
led to erroneous conclusions. Therefore, those speculations do not apply to
the observed behavior, like a large apparent pyroelectric coefficient. In
particular, the build-in bias voltage due to polarization inhomogeneity,
that has been calculated in \cite{Alpay03}, should be {\em exactly zero}, $%
4\pi \int dz\left[ P\left( z\right) -\bar{P}\right] =0,$ Eq.~(\ref{eq:field}%
). In fact, in graded samples there may be a bias voltage build-up due to a
charge trapping, etc., and this could be related to the measured anomalous
pyroelectric properties of these films, see e.g. a discussion in Ref.\cite
{Brazier99}. It is worth mentioning that dipoles, introduced by interfaces,
are likely to be important in ferroelectric superstructures, where they can
affect an electric response of the structures\cite{Superstr}{\bf .}

We thank A. Kholkin for stimulating discussions.

\end{document}